# RISK ASSESSMENT ALGORITHM IN WIRELESS SENSOR NETWORKS USING BETA DISTRIBUTION

Mohammad Momani[1], Maen Takruri[2] and Rami Al-Hmouz[3]

[1]University of Technology, Sydney, Australia
Mohammad.Momani@uts.edu.au
[2]American University of Ras Al Khaimah, United Arab Emirates
matakruri@yahoo.com
[3]King Abdulaziz University, Saudi Arabia
ralhmouz@gmail.com

## ABSTRACT

*This paper introduces the Beta distribution as a novel technique to weight direct and indirect trust and assessing the risk in wireless sensor networks. This paper also reviews the trust factors, which play a major role in building trust in wireless sensor networks and explains the dynamic aspects of trust. This is an extension of a previous work done by the authors using a new approach to assess risk. Simulation results related to the previous work and to the new approach introduced in this paper are also presented for easy comparison.*

## 1. INTRODUCTION

Trust establishment is a prerequisite for any network to function, that is, establishing trust between nodes is the first step in building the actual Wireless Sensor Network (WSN), as the creation, operation, management and survival of the WSN are dependent upon the cooperative and trusting nature of its nodes. Modelling trust requires a thorough understanding of the dynamic aspects of trust and the factors that affect trust [1]. So, in this paper, a new risk assessment algorithm for establishing trust in WSNs using Beta distribution is presented after introducing the trust factors and the main aspects of trust. This algorithm is adopting new techniques unique to WSNs. More details about the algorithm can be found in [1], which is omitted from this paper to avoid repetition.

Whenever a node in a WSN decides on whether or not to communicate with other nodes, it has to assess the other nodes' trustworthiness. Trust-modelling represents the trustworthiness of each node in the opinion of another node; thus each node associates a trust value with every other node [2], and, based on that trust value, a risk value required from the node to finish a job can be calculated. Section 2 in this paper introduces the trust factors, followed by dynamic aspects of trust discussed in section 3. Risk assessment algorithm using Beta distribution is discusses in section 4. Simulation results are presented in section 5 and section 6 concludes the paper.

## 2. TRUST FACTORS

As discussed in our previous work presented in [1], the main factors that affect the trust evaluation process of one node about other nodes are:

- Direct interactions

- Indirect interactions
- Reputation
- Risk

Direct interactions are based on self-experience or observations by one node of the other node's behaviour. A watchdog mechanism is required on each node to monitor the behaviour of other nodes in the surrounding.

Indirect interactions are based on recommendations from other trusted nodes in the surrounding area about that other node. A propagation mechanism is required to propagate the recommendations.

Reputation represents the past behaviour of a node in the absence of direct experience or recommendations.

Risk refers to the amount of risk the node is ready to take in the case of forming trust with new or unknown nodes, or in case the trust value between nodes is less than a required trust value to finish a job.

Every node in the network keeps a trust table for all the surrounding nodes it interacts with, which can include the direct trust with each node, the indirect trust, the update on both trusts, the total or combined trust and eventually the risk associated with each node.

## 3. DYNAMIC ASPECTS OF TRUST

One of the trust characteristics is dynamism, that is, trust is dependent on time, it can increase or decrease as new evidence becomes available, so the process needs to be evaluated continuously. Dynamic aspects of trust, such as the formation, evolution, revocation and propagation of trust have been discussed in depth in [3]. A brief discussion about trust formation and trust evolution in WSNs is presented below.

### 3.1. Trust Formation

Establishing trust between nodes in WSNs is the most important dynamic aspect of trust. Trust formation in WSN is the process of establishing the initial trust between nodes and in general it involves the assessment of the two main sources for calculating trust, direct and indirect interactions. Figure 1, below, shows a general trust computational model used to calculate trust in WSNs. The reputation factor is omitted from Figure 1, because it represents the past direct and indirect trust. The dispositional trust (risk) is introduced as a third source for trust calculations. The total trust is calculated by combining both trust values, direct and indirect. Knowing the trust value will lead to the risk involved in the interaction: the lower the trust value, the higher the risk, and the higher the trust value, the lower the risk.

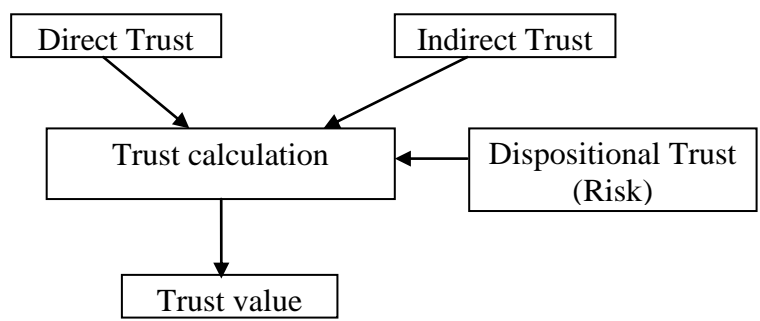

Figure 1. General trust computational model [1]

Initially, when nodes are just being deployed for the first time or when new nodes are introduced to the network, the presence of some optimistic nodes willing to take risks is required, as there is no evidence of nodes' past behaviour. Initial trust value between nodes can be assigned based on the applications and/or the deployed environment as discussed in [1].

### 3.2. Trust Evolution

The evolution process is another important dynamic aspect of trust and can be regarded as iterating the process of trust formation as additional evidence becomes available. It is the process of updating the trust level between nodes. Trust values regarding other nodes should be maintained locally and updated periodically as new evidence becomes available, and that will eventually change the risk assessment of the node. In order for nodes in a network to receive updates regarding the trusted behaviours of nodes or even threats, a mechanism for trust reporting is necessary.

As previously discussed in [1], updating trust can be achieved using the first-hand information only: direct trust, as in [4], the second-hand information only: indirect trust, as in [5, 6], the indirect positive trust only, as in [7, 8]; and/or both: direct and indirect trust, as in [9-18]. As can be seen, most systems proposed so far use both: first-hand and second-hand information.

The main issue here is how to combine these two trust sources to achieve the total trust. The traditional answer to this question is to combine them using the "weighting" approach. Some trust systems give more weights to the old experience, some give more weight to recent experience and others give more weight to direct trust rather than to indirect trust. In summary, the trust evolution process involves: updating direct trust, updating indirect trust and updating total trust, based on the updated values of direct and indirect trust.

### 4. RISK ASSESSMENT ALGORITHM

Based on the above discussion regarding trust factors and dynamic aspects of trust, a new algorithm for establishing trust in WSNs was presented as a flowchart in our work presented in [1]. The whole idea of the algorithm given in [1] is that there is a risk value associated with every job to be processed by a node, which is derived from the trust value required to do a specific task. The first thing a node will do if it has been asked to perform a task, is to compare the predefined risk value associated the task with the actual risk between the two nodes, and if the risk value is less than the predefined threshold, then the task will be performed, otherwise the task will be declined unless the node is ready to take that risk. The algorithm is just comparing risk values and combining direct trust and indirect trust to achieve the total trust and eventually calculate the actual risk associated, that is, it does not calculate the direct or indirect trust, but the node itself does that. A more detailed illustration of the algorithm follows.

It is assumed that there is a required trust value (T) associated with each job to be processed by a node and eventually a risk value is derived from that trust value. The trust value (T) is then tested against the sources of trust, the direct trust value (A), the indirect trust value (B), and the combined trust value (C) and at the same time calculates the risk value (R). If any combination of these values is greater than or equal to the required trust value, that is, the risk value is less than or equals to the predefined risk value (threshold), then the job will be processed, otherwise it will be declined unless the node is ready to take the risk associated with that job.

It is assumed that nodes are capable of calculating (A), (B) and (R) using their own criteria, be this as presented in [19] using weight and situational trust or as presented in [20] using weight and rating or any other criteria specific to a node. The challenge here, as discussed before, is how to calculate the combined trust value (C) as it represents a data fusion. In the following section a new statistical approach is presented to weight the direct and indirect trust.

## 4.1. Combined Trust

Combining direct trust and indirect trust to obtain the total trust of one node on the others was and still is the issue for many researchers. Most of the models, which use the weighting approach, assign different weights to each trust type without describing the methodology behind their assignments. Here, a new methodology using the Beta distribution is introduced to weight different trust components - direct trust value (A) and indirect trust value (B) - and eventually to calculate the total trust. The reason behind using the Beta distribution is that, it is being used widely in risk and decision analysis, due to its flexibility, and also it can be estimated very easily [8, 10].

The density function for a beta random variable X on domain [0, 1] is given by equation (1):

$$f(\alpha,\beta) = \frac{\Gamma(\alpha+\beta)}{\Gamma(\alpha)\Gamma(\beta)} x^{\alpha-1}(1-x)^{\beta-1} \tag{1}$$

$p$ is the probability that the event occurs

$$p \sim Beta(\alpha,\beta) \tag{2}$$

Substituting equation (2) in equation (1) will result in the following density function given in (3):

$$f(p) = \frac{\Gamma(\alpha+\beta)}{\Gamma(\alpha)\Gamma(\beta)} p^{\alpha-1}(1-p)^{\beta-1} \tag{3}$$

The expected value µ is given in equation (4):

$$E(p) = \frac{\alpha}{\alpha+\beta} \tag{4}$$

and the variance $\sigma^2$ is given in equation (5):

$$V(p) = \frac{(\alpha*\beta)}{(\alpha+\beta+1)*(\alpha+\beta)^2} \tag{5}$$

It is assumed that trust value (A) provides the prior estimate for $p$, as given in equation (6):

$$E_A(p) = \frac{\alpha_A}{\alpha_A+\beta_A} = \hat{P}_A \tag{6}$$

and the variance value, as given in equation (7):

$$V_A(p) = \sigma_A^2 \tag{7}$$

From the above illustration in equations (6) and (7), $\alpha_A$ and $\beta_A$ can be represented as in equations (8) and (9) respectively:

$$\alpha_A = \hat{P}_A \left[ \frac{\hat{P}_A(1-\hat{P}_A)}{\sigma_A^2} - 1 \right] \tag{8}$$

$$\beta_A = \frac{\alpha_A*(1-\hat{P}_A)}{\hat{P}_A} \tag{9}$$

Trust value (B) provides the estimate $\hat{p}_B$ of $p$, as given in equation (10):

$$E_B(p) = \frac{\alpha_B}{\alpha_B + \beta_B} = \hat{P}_B \tag{10}$$

and the variance value as given in equation (11):

$$V_B(P) = \sigma_B^2 \tag{11}$$

From equations (10) and (11), $\alpha_B$ and $\beta_B$ can be represented as in equations (12) and (13) respectively:

$$\alpha_B = \hat{P}_B \left[ \frac{\hat{P}_B(1-\hat{P}_B)}{\sigma_B^2} - 1 \right] \tag{12}$$

$$\beta_B = \frac{\alpha_B * (1-\hat{P}_B)}{\hat{P}_B} \tag{13}$$

Assuming that trust value (A) represents the prior, then according to Bayes' theorem [21],

$$P(p|\hat{P}_B) \propto P(\hat{P}_B|p) * P(p) \tag{14}$$

where, $P(p)$ represents the prior and equals to $Beta\ (\alpha_A, \beta_A)$, and $P(\hat{P}_B|p)$ is the likelihood, which needs to be modelled.

From the above discussion, equation (14) can be rewritten as shown in equation (15):

$$P(p|\hat{P}_B) \propto p^{\alpha_B-1}(1-p)^{\beta_B-1} p^{\alpha_A-1}(1-p)^{\beta_A-1}$$

$$\propto p^{(\alpha_A+\alpha_B-1-1)}(1-p)^{(\beta_A+\beta_B)} \tag{15}$$

$$\sim Beta(\alpha_A + \alpha_B - 1,\ \beta_A + \beta_B - 1)$$

Substituting equation (15) in equation (4), the following expected value will result, as shown in equation (16):

$$E(p|\hat{P}_B) = \frac{\alpha_A + \alpha_B - 1}{\alpha_A + \alpha_B + \beta_A + \beta_B - 2} = \frac{\alpha_A + \alpha_B - 1}{K} \tag{16}$$

Let *K* in equation (16) be represented as given in equation (17):

$$K = \alpha_A + \alpha_B + \beta_A + \beta_B - 2 \tag{17}$$

So, equation (16) can be written as given in equation (18):

$$E(p|\hat{P}_B) = \frac{\alpha_A}{\alpha_A + \beta_A} * \frac{\alpha_A + \beta_A}{K} + \frac{\alpha_B}{\alpha_B + \beta_B} * \frac{(\alpha_B + \beta_B)(\alpha_B - 1)}{\alpha_B * K} \tag{18}$$

$E(p|\hat{P}_B)$ in equation (18) simply represents the combined trust value, and can be written as shown below in equation (19):

$$E(P|\hat{P}_B) = \hat{P}_A * W_A + \hat{P}_B * W_B \tag{19}$$

As can be seen from equations (18) and (19), the weights for direct trust $W_A$ and indirect trust $W_B$ are represented in equations (20) and (21) respectively:

$$W_A = \frac{\alpha_A + \beta_A}{K} \quad (20)$$

$$W_B = \frac{(\alpha_B + \beta_B)(\alpha_B - 1)}{\alpha_B * K} \quad (21)$$

If the trust values (A) and (B) are known, then it is very easy to calculate, $\alpha_A$, $\beta_A$, $\alpha_B$ and $\beta_B$, as shown in equations (8), (9), (12) and (13), and eventually $W_A$ and $W_B$ will be calculated using equations (20) and (21) respectively.

As discussed before, the total trust value is a combination of direct trust and indirect trust values. One way of combining these trust values using the probability theory has been presented here.

## 5. SIMULATION RESULTS

This section presents two example simulations conducted on two different networks using MATLAB. The first network consists of three nodes for simplicity, and to be able to verify the algorithm and show the results in tables as illustrated below. The second simulation is for a network of fifteen nodes to further verify the algorithm when the number of nodes is high, and to show the results in a graph, as described below.

### 5.1. Three Nodes Network Simulation

Figure 2 below depicts a network topology of three nodes. Trust values between nodes are asymmetric, which means that the trust value from node (1) to node (2) is different from the trust value from node (2) to node (1). Table 1 below shows the complete direct trust values (A) between nodes, the required trust values (T) between nodes to perform the task, the indirect trust values (B) from the surrounding nodes, the combined trust values (C) and the risk values (R).

As can be seen from Table 1 below, the diagonal values for T and R are always zeros, which reflects the required trust value and the risk value associated with the node to itself; that is, there is no required trust value or risk associated with a node when it is doing a job for itself. The same is valid for A, B and C, which have the diagonal values set to ones, which means the node blindly trusts itself; all the values in the diagonals in all tables are from the node to itself and they do not participate in the calculations. The risk value (R) of (0) in the diagonal means the risk from the node to itself as mentioned before, and the other (0) values, not in the diagonal, mean the actual trust is greater than or equal to the required trust and there is no need for any risk to be taken.

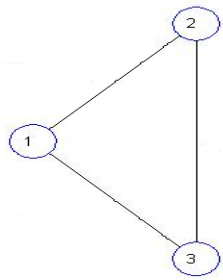

Figure 2. Network graph of three nodes

Table 1 below consists of two columns, the first column is related to trust and risk values between nodes calculated using the traditional method presented in [1] and the second column is related to trust and risk values between nodes calculated using the new method (Beta distribution) presented in this paper.

Table 1. Risk and Trust Values between nodes

| Traditional weighting method used to weight A and B | The New (Beta distribution) weighting method used to weight A and B |
|---|---|
| T = <br><br>    0    0.4546   0.7148<br>  0.7688    0    0.5383<br>  0.5846  0.2413    0<br><br>>> A<br><br>  1.0000  0.5133  0.6844<br>  0.5141  1.0000  0.1610<br>  0.4685  0.7003  1.0000<br><br>>> B<br><br>  1.0000  0.7578  0.0445<br>  0.8596  1.0000  0.5953<br>  0.4558  0.0777  1.0000<br><br>>> C<br><br>  1.0000    0    0.4693<br>    0    1.0000    0<br>  0.4928    0    1.0000<br><br>>> R<br><br>    0     0   0.2455<br>    0     0    0<br>  0.0918    0    0 | T = <br><br>    0    0.4546   0.7148<br>  0.7688    0    0.5383<br>  0.5846  0.2413    0<br><br>>> A<br><br>  1.0000  0.5133  0.6844<br>  0.5141  1.0000  0.1610<br>  0.4685  0.7003  1.0000<br><br>>> B<br><br>  1.0000  0.7578  0.0445<br>  0.8596  1.0000  0.5953<br>  0.4558  0.0777  1.0000<br><br>>> C<br><br>  1.0000    0    0.4284<br>    0    1.0000    0<br>  0.4634    0    1.0000<br><br>>> R<br><br>    0     0   0.2864<br>    0     0    0<br>  0.1212    0    0 |

In summary, the above illustrated table in this section shows the trust values between nodes and the risk associated with each node on the others connected to it in a network, the table also shows more robust values for combined trust and the associated risk using the new Beta distribution method presented in this paper.

### 5.2. Fifteen Nodes Network Simulation

To further verify the algorithm and see the effect on the trust values when many nodes exist in the surrounding area, the network should consist of a large number of nodes. Here, a simulated network consisting of fifteen nodes is presented as illustrated in Figure 3 below.

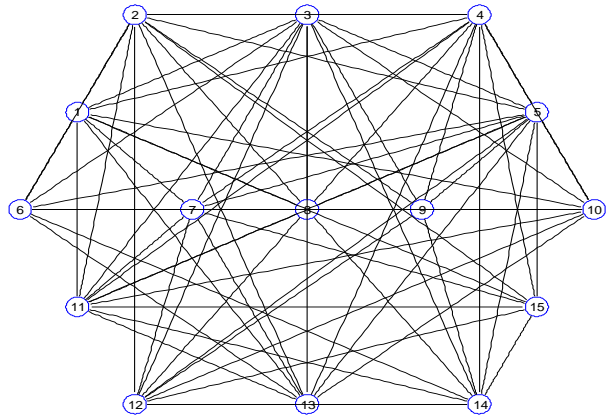

Figure 3. Network graph of fifteen nodes

Results are presented in a different way from the previous simulation, that is, risk and trust values between nodes are presented as graphs shown in figure 4 and figure 5. The graphs show the risk between every node and the other (14) nodes in the network. If there is a direct connection between any two nodes, then there will be a certain risk value, otherwise, the risk value will be equal to zero. Figure 4, shows the risk values calculated using the traditional weighting method presented in [1] while figure 5 shows the risk values calculated using the Beta distribution method presented in this paper and as can be seen from both figures the risk values calculated using Beta distribution are also more robust.

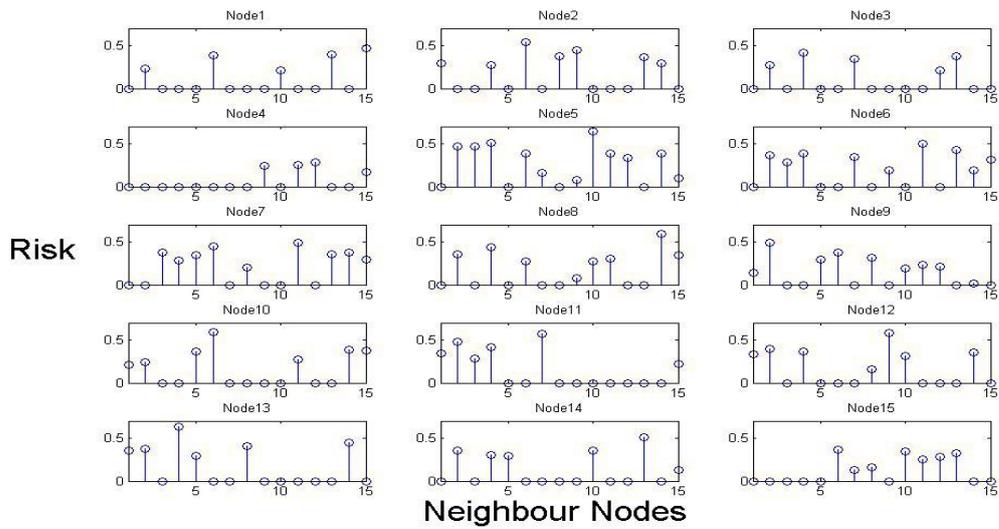

Figure 4. Risk between nodes calculated using the traditional method.

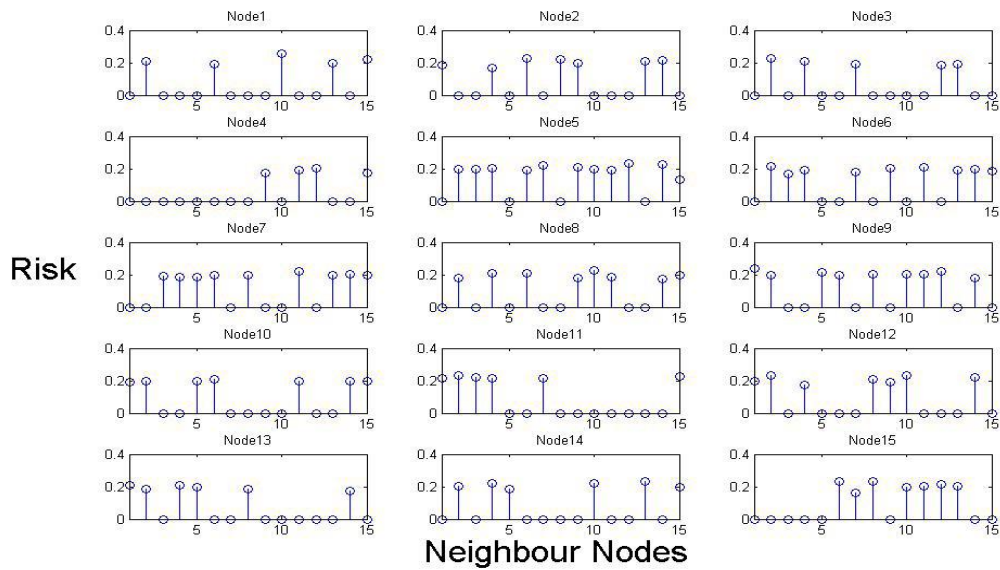

Figure 5. Risk between nodes calculated using the Beta distribution method.

## 6. CONCLUSIONS

It has been stated that modelling trust requires a thorough understanding of the dynamic aspects of trust and the factors affecting trust. This paper has introduced those two topics and briefly discussed the main factors of trust - direct trust, indirect trust, reputation and risk- with more attention given to the direct and indirect trusts as the most important factors, which can produce the other factors. A new approach to weight the direct and indirect trust using the realm of statistics (Beta distribution) for calculating trust between nodes and the risk associated with each node is also presented. Simulation results from both approaches (traditional and Beta distribution) were also presented, which simply show the trust relationship between nodes and the risk associated with them. The results have been presented in tables and in graphic format for easy observation and interpretation, showing that, the higher the trust between nodes, the lower the risk between them, and vice versa.